\documentclass{aa}
\usepackage{amssymb}
\input epsf.sty
\newif\ifAMStwofonts
\AMStwofontstrue

\def\Agata{R\' o\. za\' nska~}

\begin{document}

\title{Flare-induced fountains and buried flares in AGN}

\author{B.~Czerny, \inst{1,2} \and  R.~Goosmann\inst{3}}
\institute{$^1$~Copernicus Astronomical Center, Bartycka 18, 00-716 Warsaw, Poland\\
$^2$~associated to Observatoire de  Paris, LUTH, 92195 Meudon, France\\
$^3$~Observatoire de Paris, Section de Meudon, LUTH, Place Janssen, F-92195,
Meudon Cedex, France}

\authorrunning{B.~Czerny and R. Goosmann}
\titlerunning{Flare-induced fountains and buried flares in AGN}
\date{Received /Accepted }
\abstract{We discuss the local physical changes at the surface of an AGN
accretion disk after the onset of a magnetic flare. The X-ray irradiation
by a flare creates a hot spot at the disk surface where the plasma both
heats up and expands in the vertical direction in order to regain the
hydrostatic equilibrium. Assuming
that the magnetic loop causing the flare is anchored deeply within the
disk interior, we derive analytical estimates for the vertical
dimension $H_{hot}$ and the optical depth $\tau_{es}$ of the heated
atmosphere as a function of the position within the spot.
We perform computations for various values of the accretion rate
$\dot m$, the fraction $f_{cor}$ of radiation dissipated within the
disk corona, and the covering factor $f_{cover}$ of the disk surface
with flare-illuminated patches. It turns out that generally we
can distinguish three
characteristic radial zones within the disk showing a qualitatively
different behavior
of the heated material. In the innermost regions of the disk (inner zone)
the expansion of the disk material is restricted by strong gravitational
forces. Further out, the flare source, initially above the disk, soon
becomes embedded by the expanding disk atmosphere. At these intermediate
disk radii (middle zone) the material is optically thick thus
greatly modifying the
observed radiation by multiple Compton scattering. We show exemplary
spectra models obtained from Monte Carlo simulations illustrating the
trends. In the outermost
regions of the disk (outer zone) the expanding material is optically thin
and its influence
on the observed spectra is smaller but pressure gradients
in radial directions should cause the development of a fountain-like
dynamical structure
around the flare source. We discuss the observational consequences of our
results.

}
\maketitle

\section{Introduction}

The X-ray spectra of active galactic nuclei (AGN)
form mostly due to the Comptonization of the soft photons by the hot
optically thin plasma. Since the accretion flow, at least in bright
AGN, proceeds predominantly in the form of a cool, optically thick
accretion disk, the origin of this hot Comptonizing plasma
is not clear (see e.g. Narayan et al. 1998,
Collin et al. 2001, Czerny 2003).

One of the attractive scenarios was formulated using the
analogy with the solar corona activity: the surface of the accretion disk
may be covered with emerging magnetic loops providing natural
acceleration sites for the plasma (Galeev et al. 1979; Haardt et al. 1994).
Such a view is supported
for example by the numerical studies of the development of
magnetorotational instability inside the disk (Miller \& Stone 2000).
If allowed, large
magnetic loops emerge above the disk surface due to the buoyancy,
finally leading to a flare. The stochastic nature of the X-ray variability 
pattern
(Czerny \& Lehto 1997) is also consistent with the idea of random coronal
flares.

The energy dissipation of reconnecting magnetic loops and
subsequent efficient production of the X-ray radiation affects the disk
surface by illumination.

The irradiation of the cold (disk) surface and the formation of the
so called reflected component was studied in several papers.
Early works used the approach of the constant density medium to the disk
structure
(Basko et al. 1974; Lightman \&
White 1988, George \& Fabian 1991, \. Zycki \& Czerny 1994) while
the later papers considered also the disk vertical structure
(Raymond 1993; Ko \& Kallman 1994,
Nayakshin et al. 2000, \Agata et al. 2002).
Some of the recent papers addressed the specific issue
of the flares above the disk surface
(e.g. Ballantyne et al. 2001, Nayakshin 2000, \. Zycki
\& \Agata 2001,
Merloni \& Fabian 2002).

The reflected spectrum consists of the Compton reflection hump with
superimposed atomic features like the 6.4 keV iron line and iron edge,
recognized for the first time in the observational data by Pounds et al.
(1990). These spectral features should be the key to an understanding of
the accretion flow since they are sensitive to the ionization structure
and to the motion of the reprocessing material.

Recent observational results do not seem to favor the flare model for the
hot plasma.
High quality X-ray
spectroscopy done with Chandra and XMM-Newton satellites rather supports
the view
that strongly relativistically broadened spectral features
are characteristic for high accretion rate objects (QSO, Narrow Line Seyfert 1
galaxies) while low accretion rate objects (Seyfert 1 galaxies) display
much more narrow lines although the results are not as conclusive
as we would wish (see e.g. O'Brien et al. 2003).
Simple interpretation of this trend favors a model
with an inner hot flow. Flares may, or may not, develop only in
high accretion rate objects.

However, the irradiation of the disk surface does not only lead to a change
of the ionization state of the disk material. The heated plasma also
expands. In the present paper we concentrate on the surprising
consequences of such
an expansion of the disk surface layers under the influence
of a sudden strong flare.

\section{The model}
\label{sect:model}

Irradiation of the cold optically thick disk in hydrostatic
equilibrium by a source of intense X-ray emission leads
to the formation of a hot strongly ionized surface layer (e.g.
Nayakshin et al. 2000, Ballantyne et al. 2001,
\Agata et al. 2002, Collin et al. 2003).
The transition between the upper zone, roughly being at the Inverse
Compton temperature, and the lower much colder disk body is
very rapid if the spectrum of the incident radiation is rather hard
and the metalicity of the medium is not much lower than the solar metalicity
(e.g. El-Khoury \& Wickramashinhge 1999).
The phenomenon is related to the thermal instability
of the irradiated medium considered by Field (1965), Buff \& McCray
(1974), Krolik et al. (1981).

In the following we consider the effect of the disk irradiation by a single
flare using very simple formulae based on the works of Begelman et al. (1983),
\Agata \& Czerny (1996), Nayakshin (2000) and  \Agata et al. (2002).

\subsection{Flare at a fixed radius}
\label{sect:loc}

We assume a flare located above an accretion disk around a
black hole with the mass $M$. The flare is located above a
point in the disk at a radius $r$. The disk temperature at this
location in absence of irradiation is given by $T_{disk}$,
and the disk internal dissipation provides the energy flux
\begin{equation}
F_{disk} = \sigma T_{disk}^4,
\label{eq:Td}
\end{equation}
where $\sigma$ is the Stefan-Boltzmann constant.

The flare is expected to form due to the emergence of a
magnetic loop from the disk interior. Therefore, the height of the flare
is expected to be at the pressure scale height above the disk surface
(Galeev et al., 1979; Haardt et al. 1974;
Nayakshin \& Kazanas 2001).
Since the geometrical thickness of the accretion disk is also
roughly equal to the pressure scale height we further assume
that the disk thickness is equal to $h$ and the flare is located
at $2 h$ from the equatorial plane, i.e at $h$ above the disk
surface.

The flare luminosity is parameterized by the ratio $f_0$ of the
illuminating flux hitting the disk surface directly below the flare to the flux dissipated
within the disk interior.

We now consider the conditions at the disk surface as a function
of the projected distance from the flare location, $d$.

The incident flux from the flare decreases with distance
\begin{equation}
F_{inc} = f_0 F_{disk}{h^3 \over (d^2 + h^2)^{3/2}}.
\end{equation}
The local Inverse Compton temperature of radiation due to both the disk emission
and the incident hard X-rays can be estimated as
\begin{equation}
T_{IC} = { T_{IC}^{pl} F_{inc} + T_{disk} F_{disk}\over F_{inc} + F_{disk} },
\label{eq:tic}
\end{equation}
where $ T_{IC}^{pl}$ is the Inverse Compton temperature of the flare radiation
given by its spectral shape.
We assume that the temperature of the heated disk surface layers reaches
this value. The rapid transition between this high temperature and
the bottom cold layers takes place at the specific value of the
ionization parameter, $\Xi$, equal to $\Xi_{bot}$,
which scales with the value of the Inverse Compton temperature as
\begin{equation}
\Xi_{bot} = 1.22 (T_{IC}/10^8 {\rm K})^{-3/2},
\end{equation}
as discussed by Begelman et al. (1983).

The geometrical thickness $H_{hot}$ of the heated zone is determined by the
pressure scale height, given by its temperature and the local gravitational
field
\begin{equation}
H_{hot} = \left({k T_{IC} r^3 \over m_H G M }\right)^{1/2}
\label{eq:Hhot}
\end{equation}
where $k$ is the Boltzmann constant.

The optical depth of the heated zone is also specified by the
value of the ionization parameter $\Xi_{bot}$:
\begin{equation}
\tau_{es} = {F_{inc} \kappa_{es} r^3 \over  \Xi_{bot} c G M (H_{hot}+h)},
\end{equation}
where we assume that the medium is fully ionized and the opacity is
dominated by electron scattering, $\kappa_{es}$.

This set of equations allows us to determine $H_{hot}$ and $\tau_{es}$
as a function of $d$ if $M, r, h, T_{disk}, f_0, T^{pl}_{IC}$ are given.

\subsection{Global scaling}
\label{sect:glob}

The parameterization used in the previous section is convenient
for comparison with the result of detailed numerical computations
performed at some disk radius for a specific disk model. However,
we also introduce another, more global parameterization, which allows
to express some of the input parameters used in the previous section
by the global parameters: the total accretion rate in Eddington units,
$\dot m$, the fraction of the energy dissipated in the corona,
$f_{cor}$, and the covering factor of the disk surface, $f_{cover}$,
with hot spots. With these three parameters, we can replace three
of the parameters ($h, T_{disk}$ and $f_0$) introduced in
Section~\ref{sect:loc}.

The value of $\dot m$ specifies the accretion rate in physical units
by the relation
\begin{equation}
\dot M = 1.38 \times 10^{18} ({M \over M_{\odot} }) \dot m ~~ [{\rm g s}^{-1}].
\end{equation}

 The flux dissipated by the disk at a given radius is determined
as
\begin{equation}
F_{disk} = (1 - f_{cor}){3 G M \dot M \over 8 \pi r^3}{\cal F}(r); ~{\cal F}(r) = 1 - \sqrt{3R_{Schw} \over r}
\end{equation}
which gives the disk temperature from Eq.~\ref{eq:Td}.
Here the factor ${\cal F}(r)$ represents the standard zero-torque boundary
condition at the
marginally stable orbit.

The global accretion rate
specifies the thickness of the accretion disk which can be
approximated by
\begin{eqnarray}
h = 8 \dot m^{0.8} {\cal F}(r) \left(1 - { r\over 1400 R_{Schw} \dot m^{0.5}}\right)^{8/9} \times \nonumber\\
\left(1 -{f_{cor}\over 1 + 0.05 f_{cor}} \right) [R_{Schw}].
\end{eqnarray}
This formula is more complex than the simple formula from Shakura \&
Sunyaev (1973) for the radiation pressure dominated region since at low
accretion rates/large radii the radiation pressure contribution decreases.
The presented formula is accurate within a factor of 2 in a whole range
of disk accretion rates from $\dot m = 0.001$ to $\dot m = 1$, and radii
from the marginally stable orbit to $300 R_{Schw}$ when tested against
the numerical solutions from the code of \Agata et al. (1999). The flux
dissipated in the corona does not contribute to the disk height until
$f_{cor} \rightarrow 1$; the disk height does not tend to zero since it must
transport the angular momentum of accreting material supporting dissipation.

%
%

The ratio of the incident flux to the disk internal flux is given
by
\begin{equation}
f_0 = {f_{cor} \over (1 - f_{cor}) f_{cover}}.
\end{equation}

\begin{figure}
\epsfxsize=8.8cm
\epsfbox{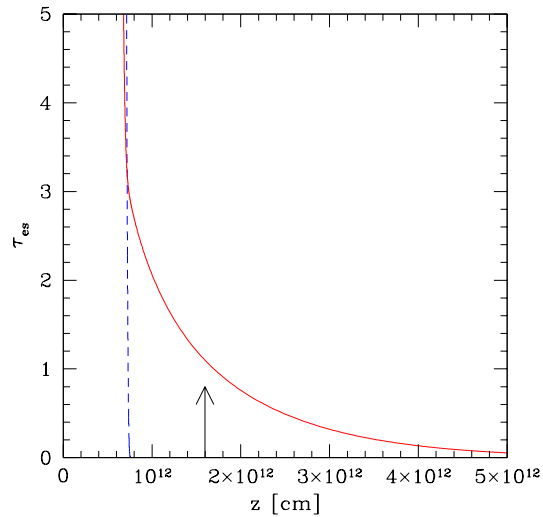}
\caption{The optical depth of the disk measured from the disk surface
as a function of the distance from the equatorial plane from the numerical
solution of the radiative transfer of Collin et al. (2003).
Dashed line - unilluminated disk,
continuous line - illuminated disk, arrow - the position of the flare at
$2h$ above the
equatorial plane. Model parameters: $M = 10^8 M_{\odot}$, $r = 9 R_{Schw}$,
$F_{inc} = 10^{15}$ erg s$^{-1}$ cm$^{-2}$, $F_{inc}/F_{disk} = 144$.}
\label{fig:numprof}
\end{figure}

We stress here that by the covering fraction, $f_{cover}$, we mean the
fraction of the disk surface which is strongly irradiated. This fraction is
specified by the mean number of flares per unit disk surface area times the
size of the hot spot (which is of the order of $\pi h^2$).
Therefore, $f_{cover}$ is
not related to the actual size of the flare itself which we assume to be
small and negligible in further considerations.

\section{Results}
\label{sect:results}

\subsection{Special case - test of the method}
\label{sect:special}

In our modeling we approximate the flare by a point-like source. We
assume that the emitted spectrum has a power law shape, with photon energies
extending from 1 eV to 100 keV. The low energy cutoff of the order of
1-10 eV is appropriate if the disk provides the soft photons for
Comptonization. The high
energy cutoff is only weakly constrained by the observational data (Gondek
et al. 1996); recent Beppo-SAX sample of Seyfert 1 galaxies gives
the best fit e-folded energy $238^{+ \infty}_{-176}$ keV
(Deluit \& Courvoisier 2003, their table 2 for
$\cos(\theta) = 0.45$). Our value of 100 keV is consistent with this limit.
The energy index $\alpha_{\rm E}$ of our
spectrum is fixed at 0.9 as it is suggested by many observations, starting
from the early
Seyfert galaxy observations done by Pounds et al. (1990). The
corresponding Compton temperature of the power law emission is
equal to $T_{IC}^{pl} = 3.27 \times 10^7 $ K.

\begin{figure*}
\parbox{\textwidth}{
\parbox{0.5\textwidth}{
\epsfxsize=0.45\textwidth
\epsfbox[18 200 600 720]{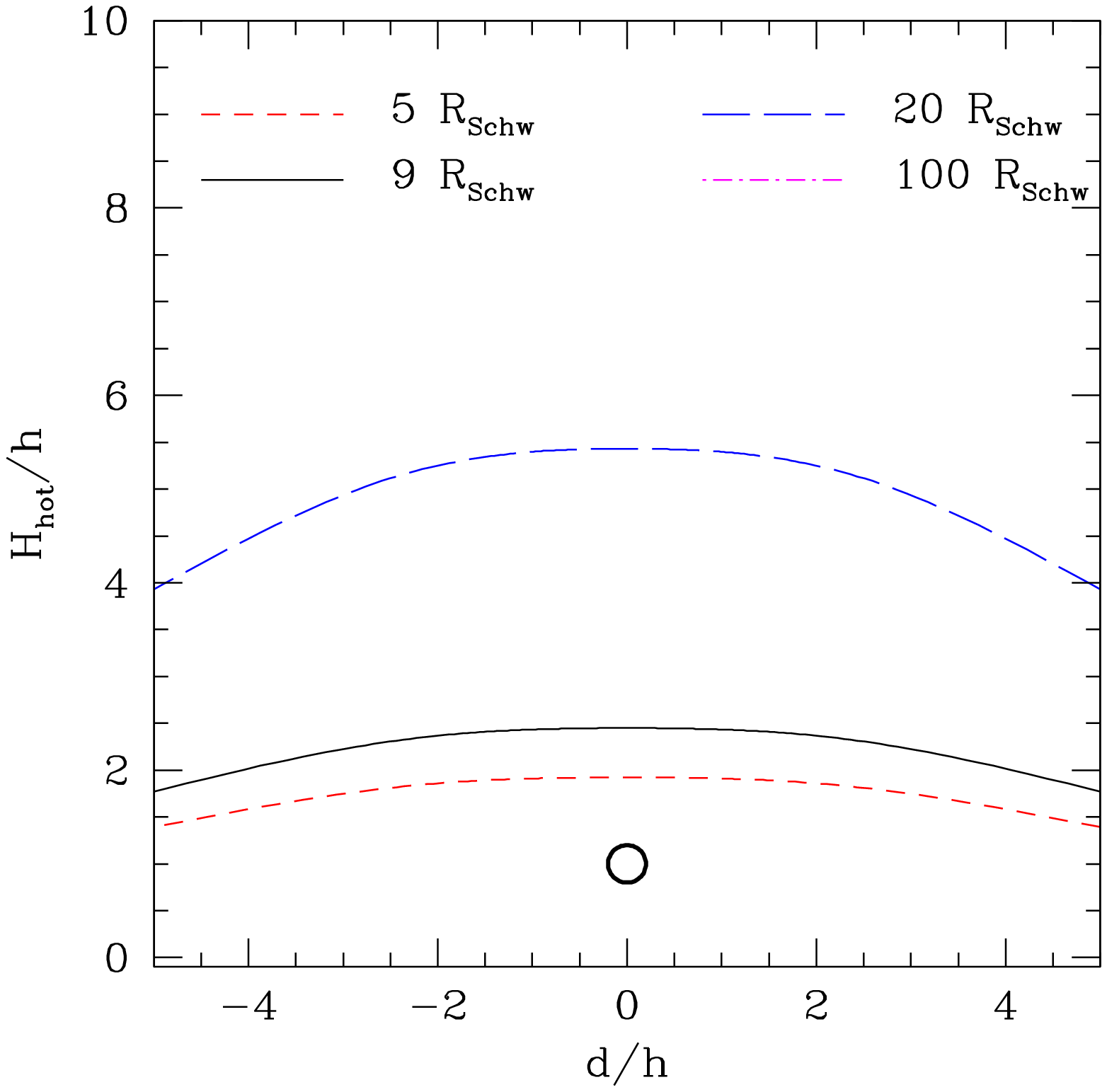}}
\hfil
\parbox{0.5\textwidth}{
\epsfxsize=0.45\textwidth
\epsfbox[18 200 600 720]{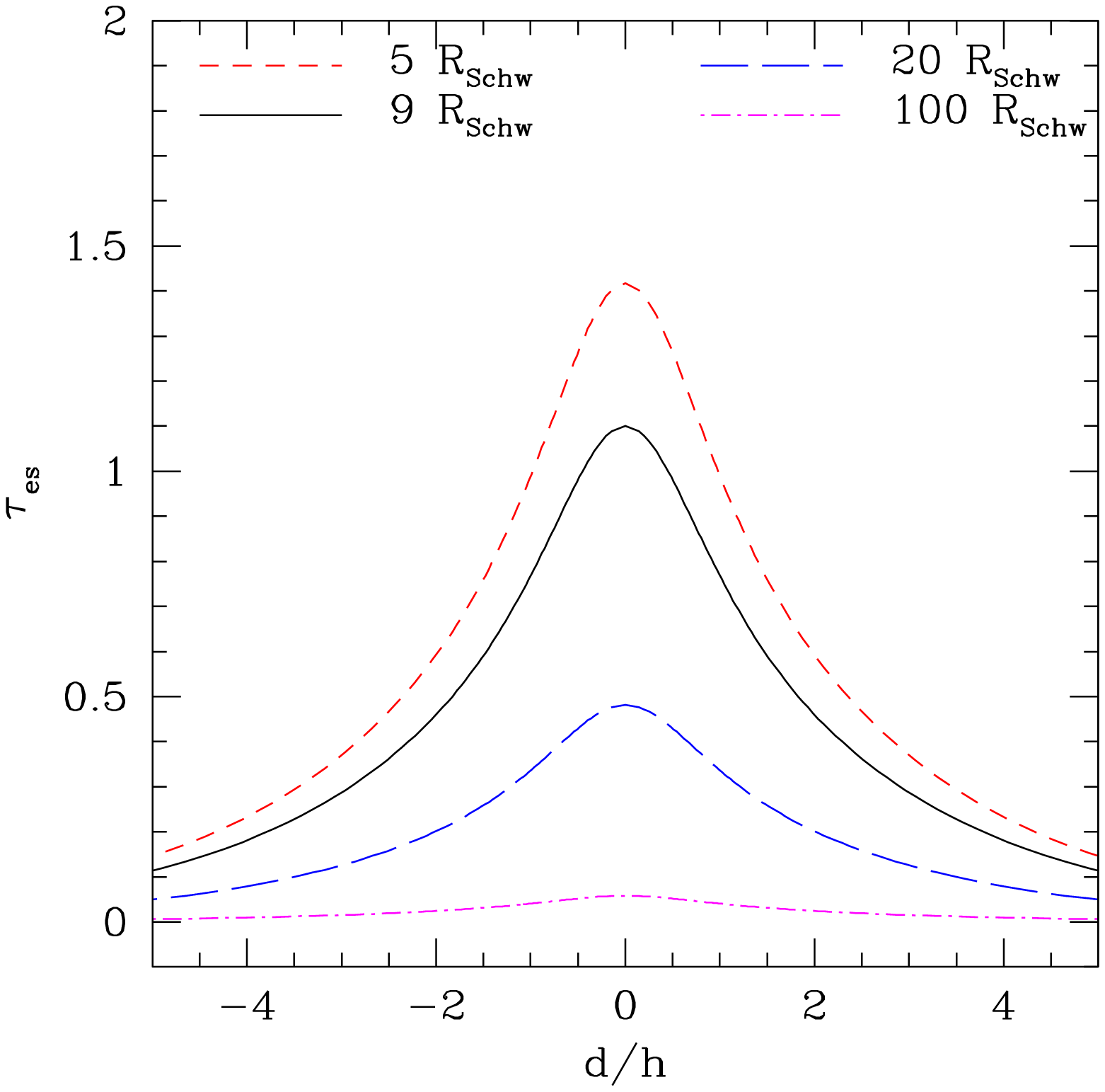}}
}
\caption{The envelope of the expanded disk atmosphere (left), and
hot region optical depth (right),  as a function
of the distance from the flare location (circle), for the special case
(parameterized as: $M = 10^8 M_{\odot}$, $ r = 9 R_{Schw}$,
$h = 8 \times 10^{11}$ cm, $T_{disk} = 1800$ K and $f_0 = 144$;
see Sect.~\ref{sect:special} ),
and for other disk radii, assuming the same
global model parameters. The shape curve for the largest radius is
beyond the scale.}
\label{fig:Ball}
\end{figure*}

We first test the model assuming parameters which were adopted in
previous numerical computations of the disk irradiation. Both Ballantyne
et al. (2001) and Collin et al. (2003) studied the radiation transfer
for a flare located above the disk around a
$10^8 M_{\odot}$ black hole, with the incident flux equal to $10^{15}$
erg s$^{-1}$ cm$^{-2}$, and the ratio of the incident radiation
flux to the disk flux equal to 144. The flare was located at 9 $R_{Schw}$
of disk radius. Since the numerical computations of the radiative transfer
by Ballantyne et al. (2001) and Collin et al. (2003) were performed
assuming the plane-parallel approximation, their results can be used to
test the model predictions directly below the flare, at $d = 0$.
We therefore adjusted the global parameters introduced in
Sect.~\ref{sect:loc} in order to reproduce these conditions.

From the analytical formulae given in Section~\ref{sect:model} we obtain
that at the location of the flare, $d = 0$, the disk heated layer
expands up to $H_{hot}/h = 2.5$, and the optical
depth of the heated zone $\tau_{es}$ is equal to 1.1. Therefore, the
heated atmosphere expands well above the original position
of the flare.

In the numerical computations of Ballantyne et al. (2001)
and Collin et al. (2003), performed until achievement of the disk
hydrostatic equilibrium, this effect was not taken into account; an
implicit assumption was made that the flare rises in order to
compensate for the expansion of the disk atmosphere. However, 
the flare supported by the disk interior
may, or may not, rise up but its evolution is caused by the dynamics
of the magnetic loop and not forced by the surrounding plasma.
We further assume that the flare remains embedded in the hot plasma,
like solar flares embedded in the solar corona. As a more accurate approach
we should in this case consider the modification of the irradiating flux
by the scattering within the hot medium. However, for simplicity we
neglect it in the present research. Therefore, our results should be
directly comparable to the numerical results.

The comparison of the
optical depth measured from the disk surface for
the non-irradiated disk and irradiated disk in hydrostatic equilibrium
is shown in Fig.~\ref{fig:numprof}. The density in the unilluminated disk
rises fast close to the surface so the profile is very steep (dashed line)
and marks well the disk surface. The density profile in illuminated model
is much more shallow (continuous line). The disk thickness, defined by the
density drop to $10^{-16}$ g cm$^{-3}$, increases from  $8.0 \times 10^{11}$
to $1.3 \times 10^{13}$ cm, so by a factor of 10, but most of the plasma
is located within the distance 2-3 larger than the original disk height.
The optical depth of the medium measured from the position of the flare
(marked by an arrow) is equal to 1.3. These numbers are not far from
our analytical estimates. This is encouraging for our simple approach:
the analytical results should be valid within a factor of a few or better.

The advantage of our model is that we can perform computations not
just below the flare, but as a function of the projected distance.

The distribution of the thickness of the Compton heated zone
as a function of the distance $d$ for this model is
shown in Fig.~\ref{fig:Ball}.

From Fig.~\ref{fig:Ball} we see that the geometrical thickness
of the heated zone decreases with $d$ rather slowly. This is due to the
fact that this quantity is determined by the Inverse Compton
temperature. Due to the large adopted ratio of the flare
to disk flux at $d=0$, the Inverse Compton temperature is
dominated by the flare contribution even at large distances. On the
other hand, the
incident flux itself decreases more rapidly and so does the optical
depth of the heated zone and its density.

The heated layer has moderate optical depth, of the order of 1 close
to the flare. Therefore, the flare is not completely shielded from
an observer by the plasma environment but the scattering is
significant and may influence the flare's appearance. We neglect this
effect in our model but we will return to this issue later on.

\subsection{General case: characteristic radii}

\begin{figure}
\epsfxsize=8.8cm
\epsfbox[70 190 400 680]{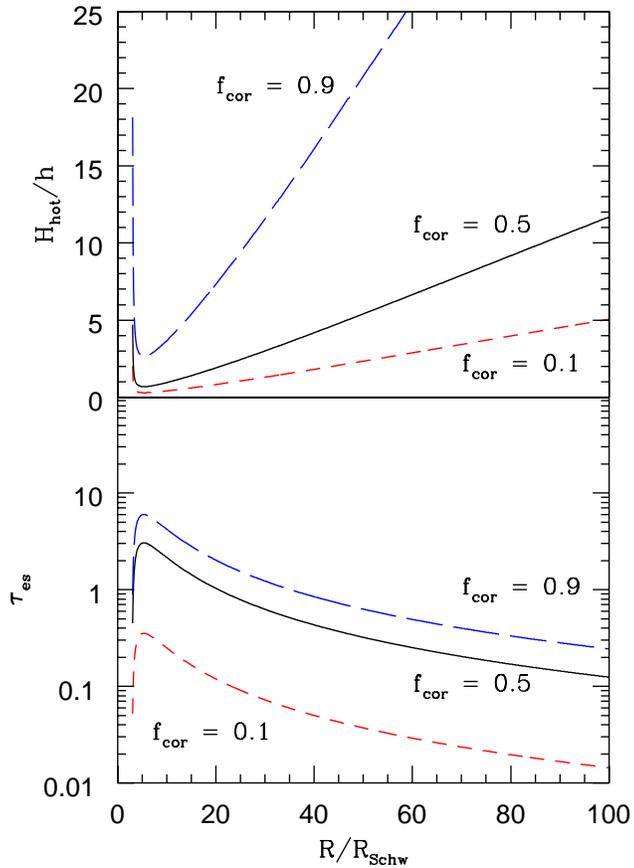}
\caption{Radial dependence of the thickness of the heated zone (in flare
height units) and the optical depth of the heated zone directly at the flare
position ($d = 0$) for three values of the fraction of energy
dissipated in the corona, $f_{cor}$. Other model parameters: $M = 10^8 M_{\odot}$,
$\dot m = 0.03$, $f_{cover} = 0.1$.}
\label{fig:glob1}
\end{figure}

In order to see the direction in which the hot medium properties
change we plot in Fig.~\ref{fig:glob1} three examples of
the radial dependence of the geometrical
thickness of the heated zone and the optical depth of the heated
zone just at the flare location, $d = 0$. 

Close to the marginally stable orbit the optical depth drops rapidly
and the disc thickness (and consequently flare height) go to zero which is 
related
to the zero-torque boundary condition used in the model. This is a property
of all presented models and we do not discuss the results in this region.

Further out, we can distinguish three types of characteristic zones:
\begin{itemize}
\item {\bf outer zone}: geometrically thick but optically thin.
In this zone the flare is embedded in the hot medium; it still
can act as a point like source although a certain fraction of the emitted
photons is scattered by the expanded plasma.
\item {\bf middle zone}: geometrically thick and optically thick. Our
simplified computations do not strictly apply to this zone but the existence
of the zone is clearly indicated; here the flare is located deep inside
the hot medium so the radiation coming out cannot be clearly seen as a
point like source.
\item {\bf inner zone}: geometrically thin; here the flare
is not at all embedded in the heated material.
\end{itemize}

The transition from the outer to the middle zone is determined by the
condition $\tau_{es} = 1$ of the heated layer. The transition
between the middle and the inner zone is defined by the condition
$H_{hot}/h = 1$. In each of these zones the material is expected
to behave differently.

\begin{figure*}
\parbox{\textwidth}{
\parbox{0.5\textwidth}{
\epsfxsize=0.45\textwidth
\epsfbox[18 200 600 720]{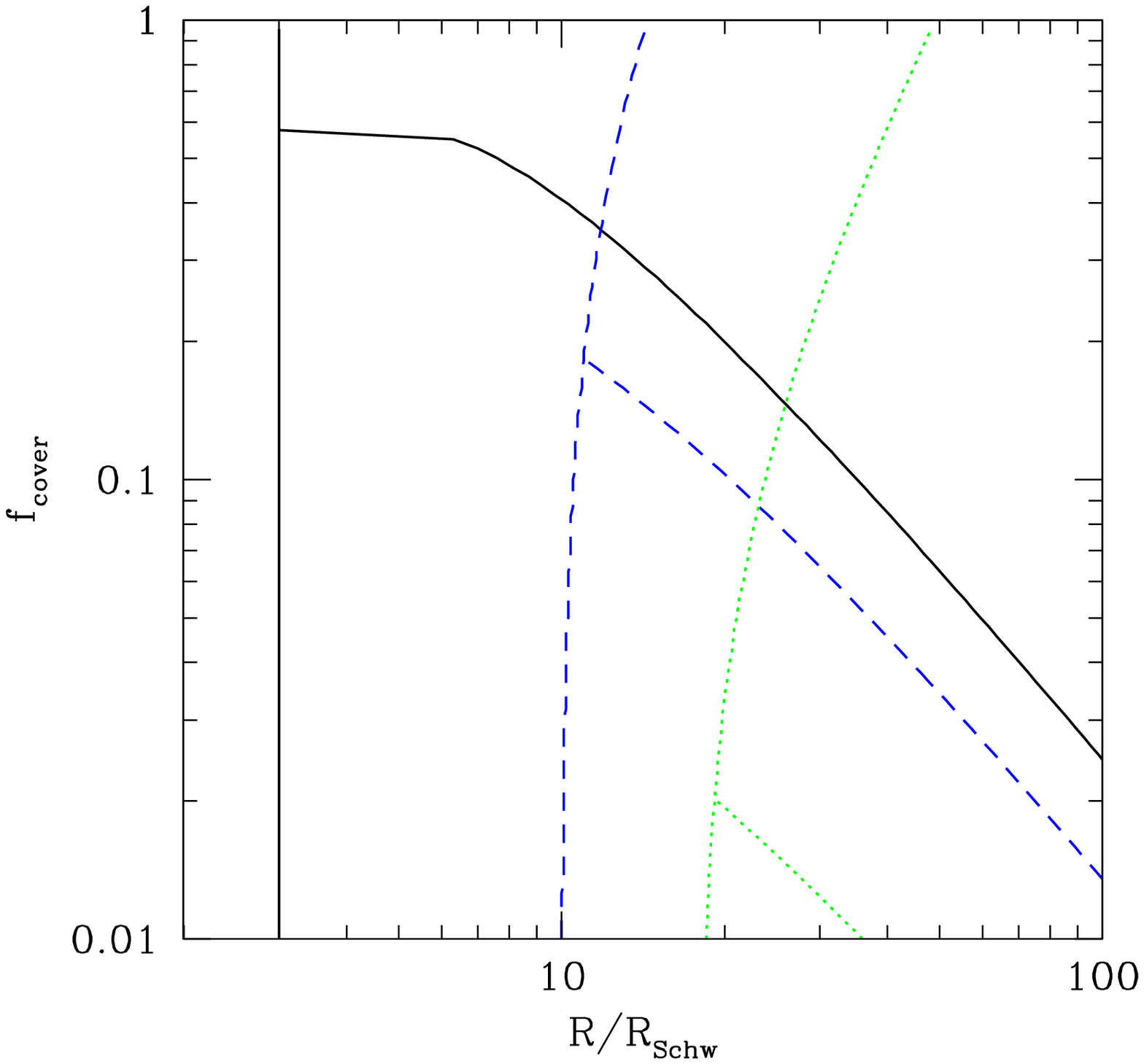}}
\hfil
\parbox{0.5\textwidth}{
\epsfxsize=0.45\textwidth
\epsfbox[18 200 600 720]{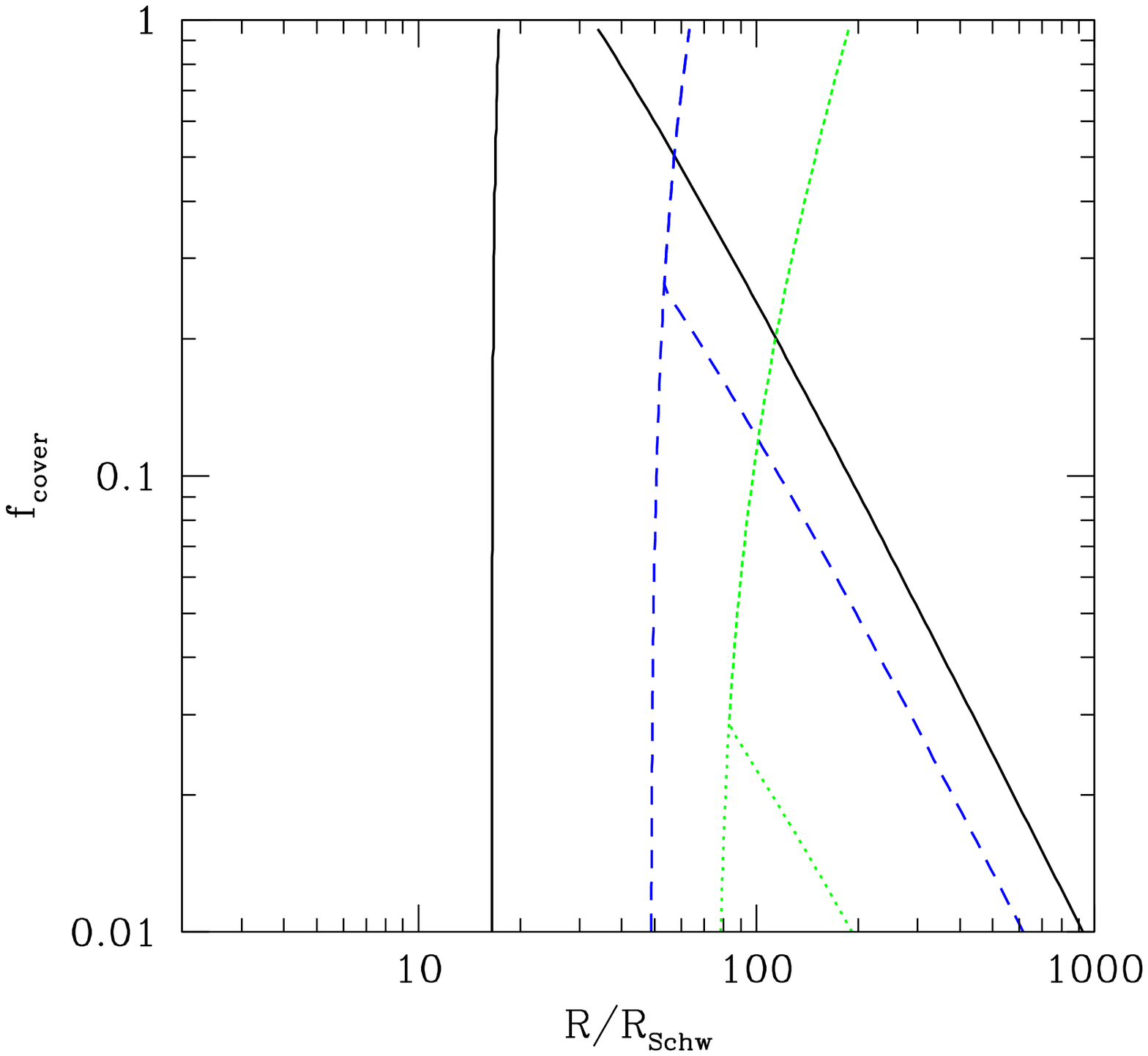}}
}
\caption{Radial extension of the three zones for $\dot m = 0.03$ (left)
and $\dot m = 0.3$ (right) as a function
of the covering factor for three values of $f_{cor}$: 0.1 (dotted line),
0.5 (dashed line) and 0.9 (continuous line). The inner zone is to the
left from the lines of a given style and the outer zone is to the right.
The middle zone exists only for small values of $f_{cover}$: it is
located between the two
lines of a given style and the lines merge for large$f_{cover}$. Generally,
the extension of the middle zone decreases with the decrease of the
coronal strength and with the
increase of the covering factor. Black hole mass: $M = 10^8 M_{\odot}$.}
\label{fig:rad}
\end{figure*}

Radial extension of the zones depends significantly on the
adopted values of the parameters, $f_{cor}$, $f_{cover}$ and $\dot m$. 
A few examples are shown
in Fig.~\ref{fig:rad}.
For a strong corona, $f_{cor} = 0.9$, the middle region is usually 
very extended
but it shrinks with an increase of the covering factor. The inner region with
bare flares shows up only at high accretion rates. For a
moderate corona,  $f_{cor} = 0.5$, the middle region is generally less extended
so for a weak corona/high
covering factor the middle region disappears and only the outer region and
an inner region remain.

However, the radial extension of the zones depends on the adopted accretion
rate.
For sources radiating closer to the Eddington rate an inner zone
always develops
(see Fig.~\ref{fig:rad}, right panel). The width of the middle zone in high accretion rate
objects is slightly narrower than in the case of low accretion objects,
if measured in logarithmic scale, and
it is systematically shifted toward larger radii, making space for
an inner zone. The basic trends with the covering
fraction and the corona strength remain the same: stronger corona (i.e.
larger $F_{cor}$) leads to broader middle zone while larger covering
factor causes it to shrink.

\subsection{The outer zone, $\tau_{es} < 1$, $H_{hot}/h > 1$}

\begin{figure*}
\parbox{\textwidth}{
\parbox{0.5\textwidth}{
\epsfxsize=0.45\textwidth
\epsfbox[18 200 600 720]{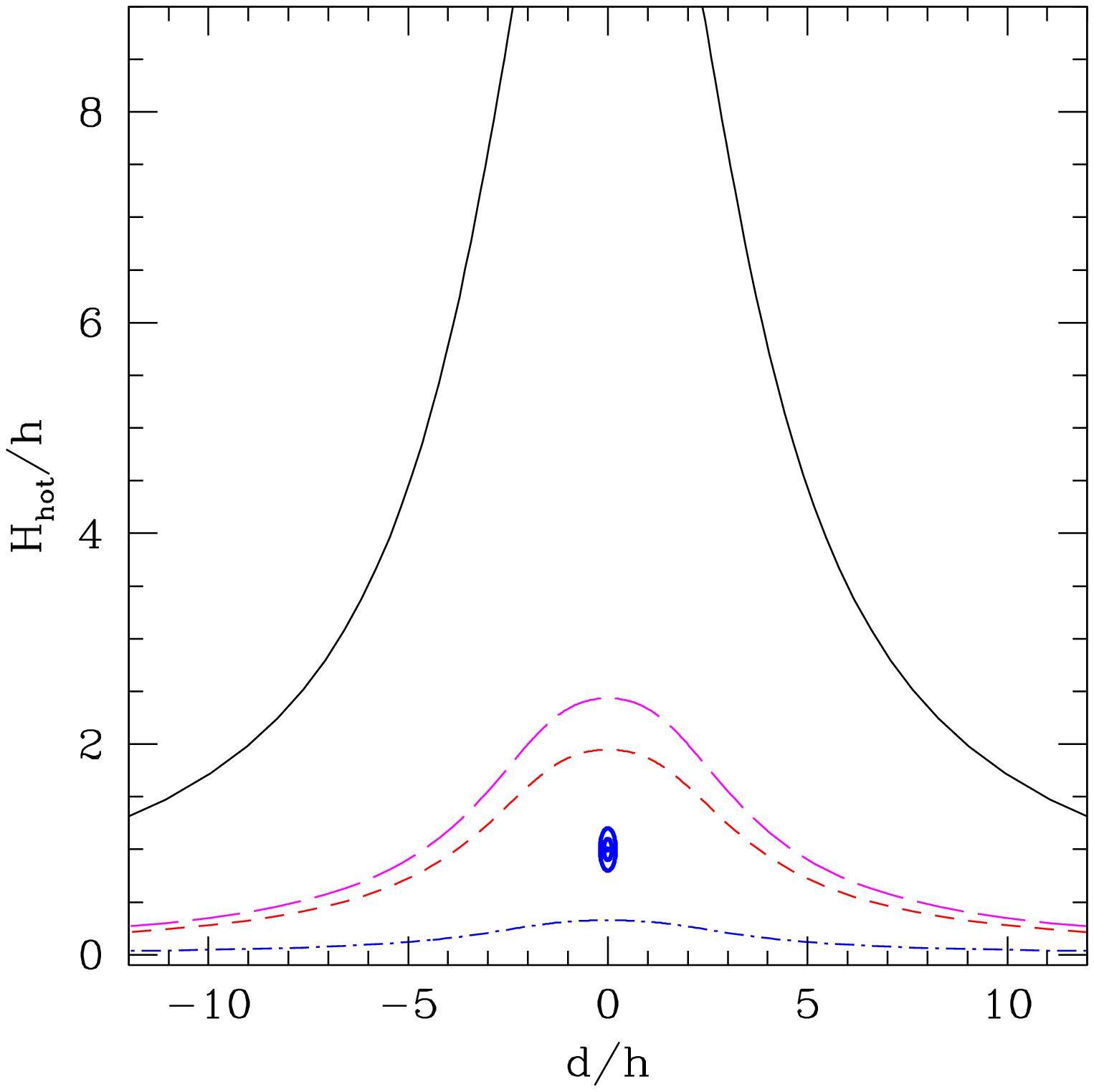}}
\hfil
\parbox{0.5\textwidth}{
\epsfxsize=0.45\textwidth
\epsfbox[18 200 600 720]{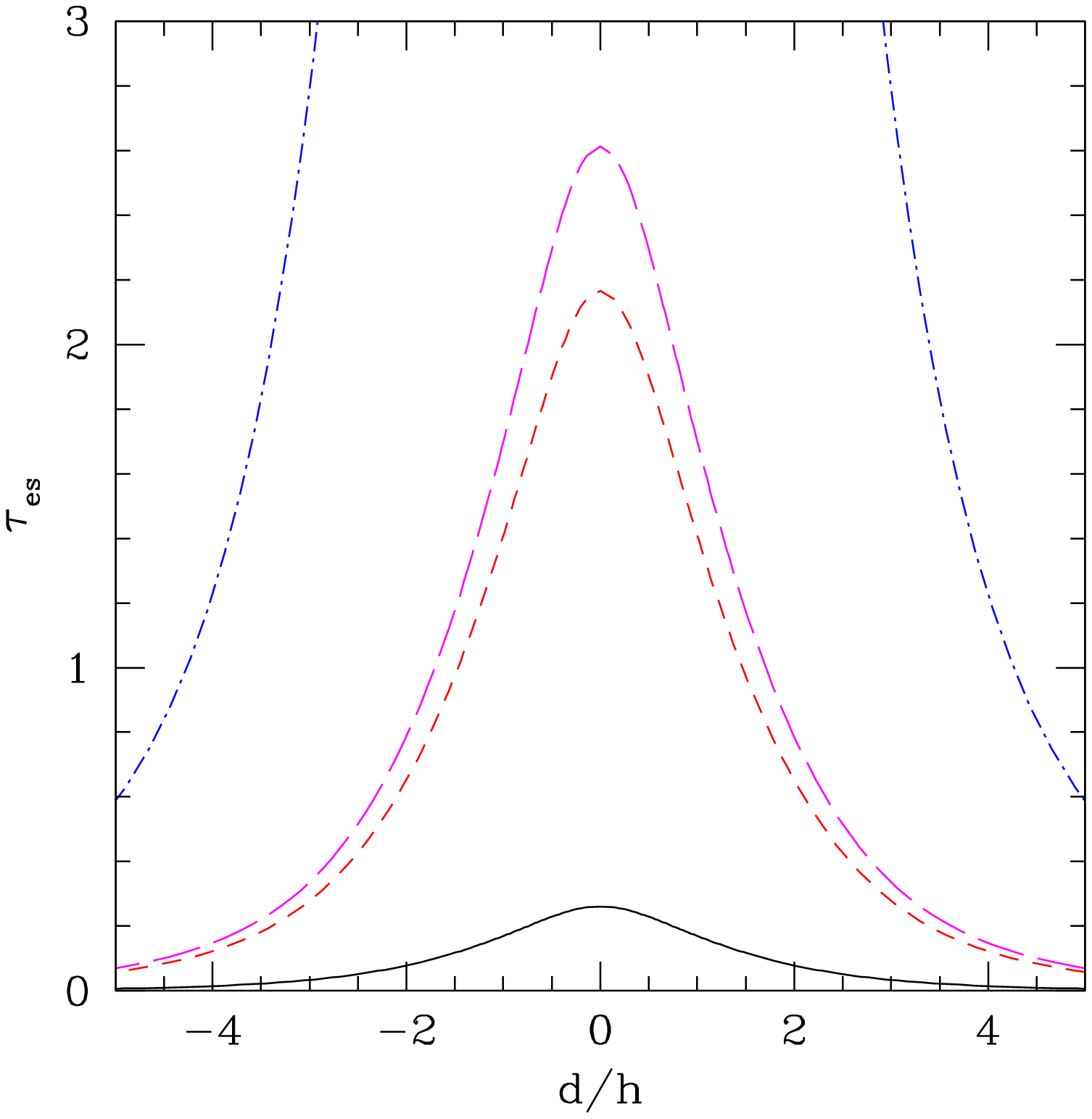}}
}
\caption{The envelope of the expanded disk atmosphere (left) as a function
of the distance from the flare location, and the
hot region optical depth (right) for four parameter sets:
$\dot m = 0.03$, $R = 100 R_{Schw}$ (example of the solution in outer zone;
solid line), $\dot m = 0.03$, $R = 20 R_{Schw}$ (example of the solution in
middle zone; short-dash line),  $\dot m = 0.3$, $R = 100 R_{Schw}$
(example of the solution in
middle zone; long-dash line),  $\dot m = 0.3$, $R = 20 R_{Schw}$
(example of the solution in
inner zone; dash-dot line).  Other parameters: $M = 10^8 M_{\odot}$,
$f_{cover} = 0.05$, $f_{cor}=0.5$.}
\label{fig:outer}
\end{figure*}

This zone is present in all solutions, for each parameter set.
Our computations are strictly applicable there. The flare is embedded
in the hot expanded medium but the medium is optically thin. Therefore,
both the point-like source is visible to the
observer as
well as the disk surface is irradiated roughly like in the absence of the
scattering medium.

An exemplary profile of the geometrical thickness and the
optical depth of the medium in the vicinity of the flare is shown in
Fig.~\ref{fig:outer}.
For a parameter set we considered
($f_{cor} = 0.5$, $f_{cover} = 0.1$, $R=100 R_{Schw}$, $\dot m = 0.03$)
the profile is less shallow than in
Fig.~\ref{fig:Ball} since the ratio of the incident to disk flux underneath the
flare is
now 10 instead of 144 so the value of the Inverse Compton temperature (see
Eq.~\ref{eq:tic}) drops more rapidly. However, the basic picture remains
the same: the hot material rises most effectively directly under the flare,
surrounding it, and less effectively further away.

\begin{figure}
\epsfxsize=8.7cm
\epsfbox[15 200 400 480]{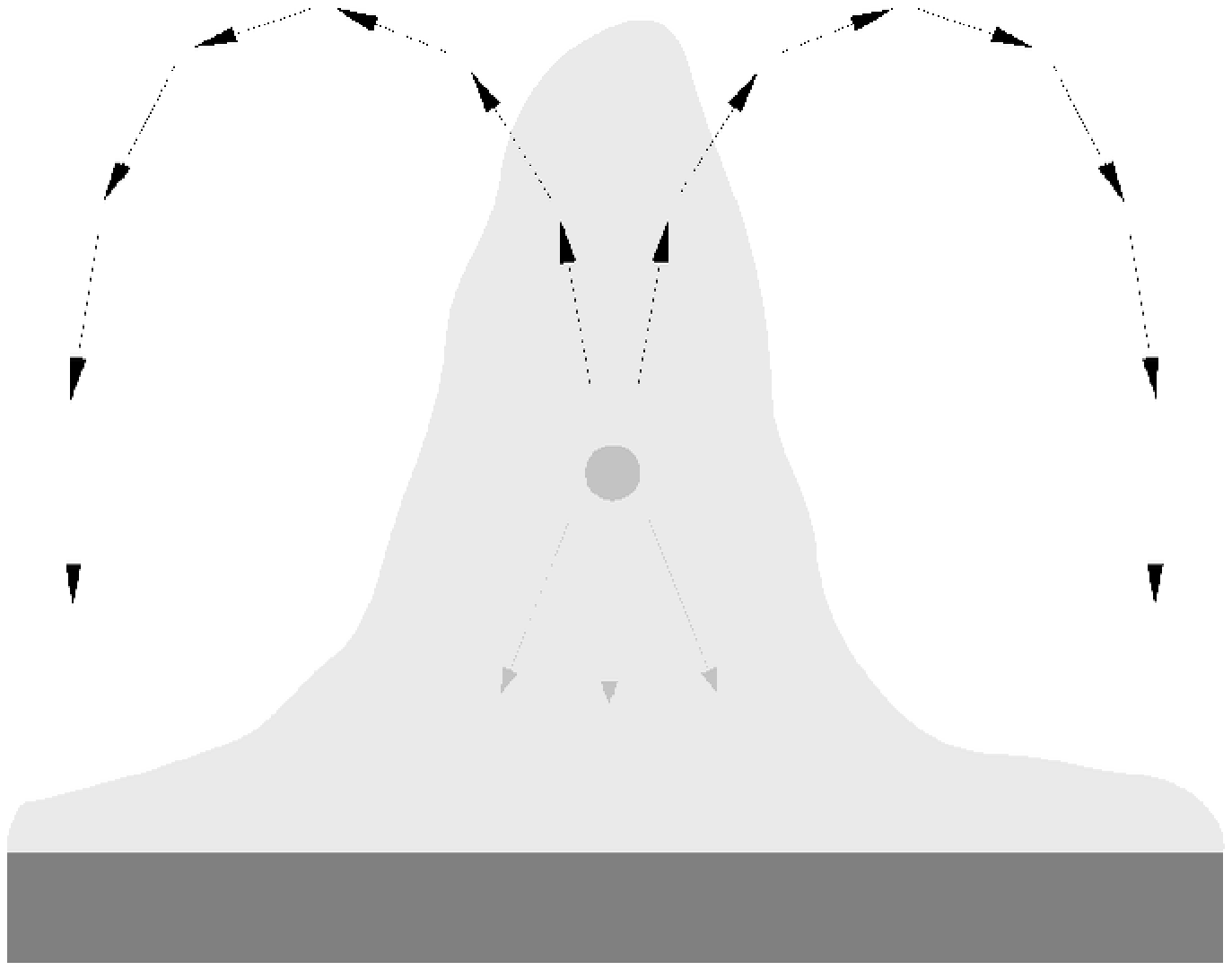}
\vskip 1 truecm
\epsfxsize=8.7cm
\epsfbox[15 200 400 480]{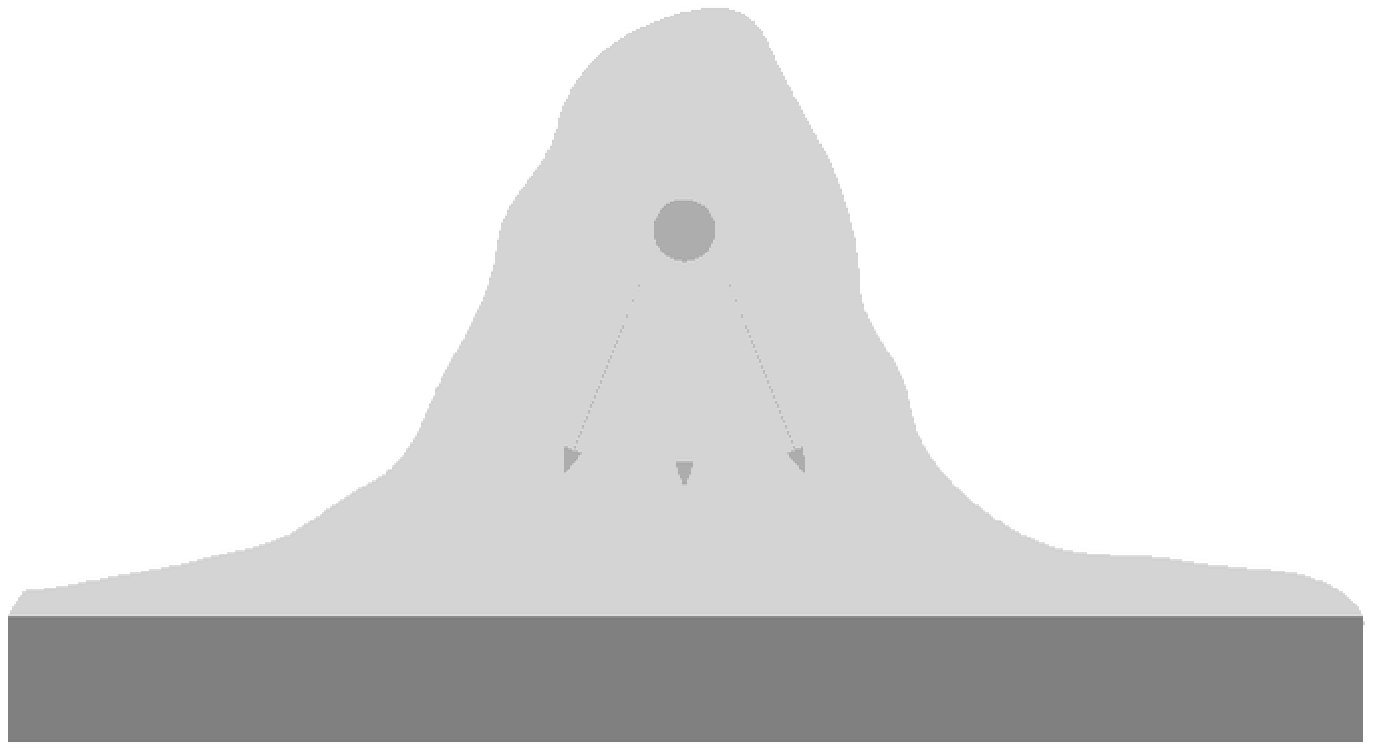}
\vskip 1 truecm
\epsfxsize=8.7cm
\epsfbox[15 200 400 480]{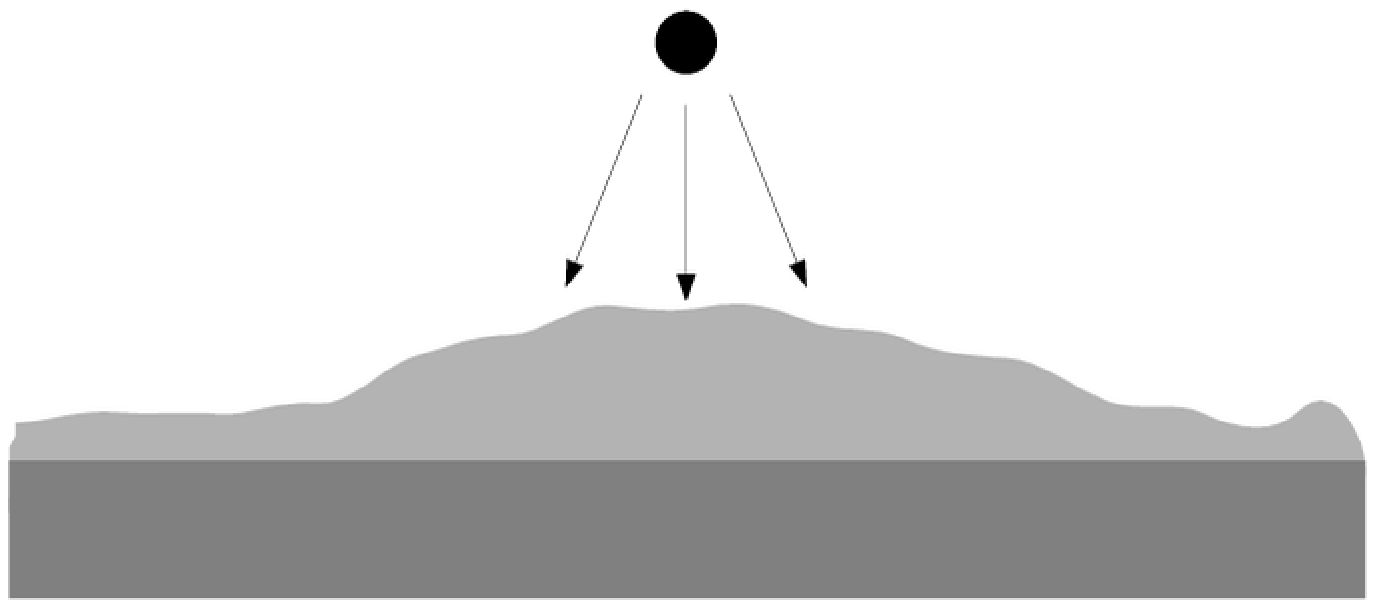}
\caption{The fountain-type flow induced by the presence of a compact
flare irradiating the disk surface in the outer zone (upper panel), the
flare embedded in the optically thick plasma in the middle zone (middle
panel), and the bare flare
in the inner zone (lower panel).}
\label{fig:draw}
\end{figure}

If the flare lasts long enough, a kind of semi-stationary picture is
expected to emerge. Radial pressure gradient should cause a side flow of
the evaporated material and its subsequent condensation at the disk surface
while a fresh material will rise up. 
This circulation creates a fountain-like phenomenon.
We envision this situation qualitatively in Fig.~\ref{fig:draw}  but the
computations of the dynamics of the plasma are beyond the scope
of the present paper.

Therefore, in the outer zone, the flare above an accretion disk shows
considerable similarity with ejective flares observed in the solar
corona. Solar flares, being the sources of the hard X-ray emission,
are also embedded in the optically thin solar corona, and accompanied
by the flow of the heated material (e.g. Raymond et al. 2003;
Manoharan \& Kundu, 2003 for recent solar observations).

The flow of the material in AGN will be additionally affected by the
disk rotation.
In a Keplerian disk, the local epi-cyclic frequency is also given by the
local Keplerian frequency so the flow pattern will be strongly twisted due to
Coriolis forces.

\subsection{The middle zone, $\tau_{es} > 1$, $H_{hot}/h > 1$}

This zone covers a major part of the disk in the sources with low accretion
rate and strong corona. The optical depth of the hot material, as computed
from our approach, varies from just above 1 in models with relatively large
covering fraction to $\sim $ 10 in models with $f_{cover} = 0.01$.

Tho examples of the solution when the optical depth is still relatively low
are presented in Fig.~\ref{fig:outer}, for parameter sets:
$R=20 R_{Schw}$, $\dot m = 0.03$ and $R=100 R_{Schw}$, $\dot m = 0.3$,
assuming in both cases $f_{cor} = 0.5$, $f_{cover} = 0.05$.

Our analysis allows us to estimate the extension of the zone. However, a
quantitative description of the processes going on there is not
possible within the adopted approach. In this zone the flare is
deeply embedded in an optically very thick plasma so any photon is
scattered many times before leaving the area. From the point of view of
an observer, a compact source is now replaced by a more extended
medium surrounding it.
The irradiation of the disk surface is
partially through the scattered radiation. The
exact computations would be particularly complex in these zone, with the
effect depending both on the total optical depth of the heated layer and
on the amount of the material between the flare itself and the cold disk
surface (i.e. on the $H_{hot}/h$ ratio).

Since in this zone also the radial pressure gradients will
develop within the heated material we also expect some systematic
circulation of the material in this zone. However, it may be less
spectacular due to the large optical depth, effectively forming
a kind of turbulent optically thick skin covering the disk surface.

\subsection{The inner zone, $H_{hot}/h < 1$}

This zone is always present in high accretion rate objects.
Here the flare is
located above the heated plasma and a fraction of the flare
emission reaches an observer directly. From this point of view,
the inner zone is similar to the outermost, optically very thin part
of the outer zone. In Fig.~\ref{fig:outer} we give a possible solution of the
inner region for the parameters $R=20 R_{Schw}$, $\dot m = 0.3$,
assuming in both cases $f_{cor} = 0.5$, $f_{cover} = 0.05$.

The optical depth of the heated layer is often high; its value is not
well estimated from our model since in the case of the large optical depth
the radiative transfer must be considered. The extension of the zone, however,
is well determined.

This region is well described by the models of Ballantyne et al. (2001)
and Collin et al. (2003) who neglected the possibility of the flare being
surrounded by the rising disk atmosphere. No spectacular
fountains are expected as the rise of the material is suppressed by
the strong gravitational field. The role of the radial pressure gradients in
this situation must be smaller, as the ratio of the radial to vertical
pressure gradient scales with $H_{hot}/d$, and consequently the radial flow
velocities are expected to be much lower than the sound speed.

\section{Discussion}
\label{sect:discussion}

The flare scenario is an attractive explanation of the origin of the
hard X-ray emission in radio quiet AGN. However, this idea mostly exists as
a simple phenomenological picture. If we want to prove its viability -
or reject the model - we must study more carefully its implications.

In the present paper we show that the development of a flare leads not
only to the irradiation of the disk surface but it strongly perturbs the
disk surface structure.

Taking into account the reaction of the disk surface to the irradiation, we
divide the disk into three zones. In the outer zone the flare is soon
embedded in the rising medium but the plasma surrounding the flare is
optically thin. In the middle zone the flare is embedded in an optically
thick plasma. In the inner zone the flare stands above the perturbed disk
surface. The middle zone does not always exist: it is broad for low covering
factor for the hot spots and/or large fraction of energy dissipated
in the form of the flare emission, and disappears for large
covering factor and/or weak flares.

This rising plasma, heated up to the local Inverse Compton temperature,
may influence the observational appearance of a flare.

\subsection{Observational consequences}

If the assumed model parameters allow the existence of the middle
region the formation of the X-ray reflected component in the source is strongly
modified. In the middle region the flare is buried deeply within the optically
thick plasma. This will have two important consequences.

\subsubsection{Formation of the reflected component}

Since the flare emission diffuses through the optically thick medium,
no reflection component forms in this region of the disk. Therefore,
the  reflection forms only in the outer as well as in the innermost
zones. The relativistic broadening  of the spectral features of this
component (mostly the iron line) allows to identify the formation
region so studying the broadening we should see two components: one,
only weakly broadened, and the second one coming from the relatively
narrow inner ring.

As an example, we can use the model parameters possibly appropriate
for the famous source MCG -6-30-15. This object  radiates most probably
at $\sim 0.1$ (Czerny et al., 2001) of the Eddington ratio. Assuming
the covering of the disk surface with hot spots $f_{cover} = 0.05$
and the corona strength $f_{cor} = 0.8$ we obtain that the inner
reflection-forming ring ends at  $12R_{Schw}$. It may explain  why the
reflection observed in this source  comes only from a relatively
narrow ring, $R_{out} = 10 R_{Schw}$ (Lee et al. 2002b), instead of
the whole disk.

If the source accretes at a low rate
and the corona is strong the inner zone
practically disappears and no reflection component forms in the inner
region of the
disk. This is for example the case for sources with $\dot m = 0.03$ when
$f_{cor} > 0.65$ and $f_{cover} < 0.5$. Our picture may therefore revive
the idea pursued by Young et al. (2001)
in the context of Cyg X-1 that the accretion disk also for low luminosity
states extends down to the marginally stable orbit. This view was
criticized by Barrio et al. (2003) at the basis of a careful
study of the reflection
component, assuming that the  incident radiation comes from above the disk
atmosphere. However, if the 'incident flux' comes from within the optically
deep surface layers, the analysis of Barrio et al. (2003) does not apply.

Some reflection also forms in the outer region: for the case considered
for MCG -6-30-15
($\dot m = 0.1$, $f_{cor} = 0.8$, $f_{cover}=0.05$) this region extends
outward from
$\sim 130 R_{Schw}$; the actual value depends significantly on the adopted
 $f_{cover}=0.05$. The amount of gravitational energy available there is
already quite low. This region should contribute a relatively narrow and
not too strong iron line component to the observed reflection spectrum.

Burying a flare inside an optically thick region is even more efficient in
suppressing the reflection than complete ionization of the illuminated surface
discussed by Nayakshin et al. (2000), or Ballantyne et al. (2001) and
Done \& Nayakshin (2001). It also allows for a natural possiblity of
a relatively rapid radial transition from the inner reflection-producing 
region (bare flares) to the middle non-reflecting region.

\subsubsection{Thermal vs. non-thermal character of the 'primary' emission}

In the context of the solar flare models, both
thermal radiative processes (e.g. Karpen et al. 1989; Gan et al. 
1991) as well as non-thermal emission (Mariska et al.
1989; Antonucci et al. 1993) were considered. Both types of processes
are probably
relevant also in the case of magnetic flares above AGN accretion
disks and accretion disks in galactic sources
(e.g. Coppi 1992, Gierli\' nski et al. 1999, Wardzi\' nski \& Zdziarski 2001,
Vaughan et al. 2002).
Together they form the (customarily named) 'primary component' of the
radiation spectrum, but their
relative importance seems to depend on the accretion rate.
Comptonization by mostly thermal electrons ($T \sim 100$ keV is
favored for low $\dot m $ objects while a significant non-thermal electron
population is likely to be present in high $\dot m $ objects.

Those trends with the accretion rate may be possibly explained within the frame
of our model.

High $\dot m $ objects in our picture are characterized by the extended
inner zone with bare flares. In this case we might expect a significant
fraction of non-thermal electrons.

However, if a flare is embedded in a surrounding plasma, several
additional processes are expected. There will be a significant down-scattering
of the photons at energies around $\sim 100 $ keV. Also cooling and
thermalization of the electrons may be different in such surroundings.
As a result, the emission coming from the middle region may look
much more 'thermal' than the emission from the inner zone.

To illustrate the trend, we calculate the observed spectrum from a buried flare
assuming that the surrounding material forms a spherical cocoon
around the flare location. The plasma is assumed to be completely ionized, and
parameterized with the optical depth. The point-like flare source of radiation
is modeled as a power law with the energy index
$\alpha_{\rm E} = 0.9$ (photon index $\Gamma = 1.9$) extending from 0.1 keV
up to 1 MeV. We choose such an extreme case to illustrate better the
spectral shape trends. The temperature of the medium is set at the value of
the inverse Compton temperature for such an incident spectrum ($T_{IC} \approx 2 \times 10^8$ K).
The setup is basically similar to models describing
Comptonization by cold and/or hot electrons considered in the past in many
papers (e.g. Pozdniakov et al. 1979, Sunyaev \& Titarchuk 1980, G\' orecki
\& Wilczewski 1984; for more recent developments see Titarchuk \& Hua 1995,
Hua \& Titarchuk 1995, and references in these articles).

\begin{figure}
\epsfxsize=8.8cm
\epsfbox{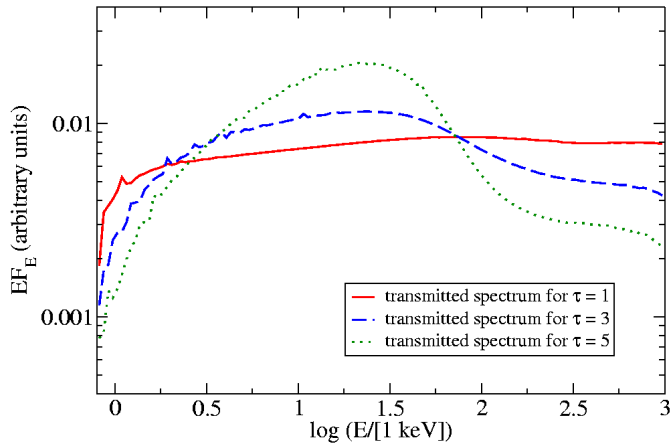}
\caption{The spectra of buried flares for
a sequence of values of the optical depth in the surrounding plasma obtained
with the Monte Carlo code {\sc noar}.
Plasma temperature - 18 keV, intrinsic flare spectrum of a power law shape,
$\alpha_{\rm E}=1.0$, $E_{\rm min} = 0.1$ keV, $E_{\rm max} = 1 $ MeV}
\label{fig:MonteCarlo}
\end{figure}

The results obtained with the Monte Carlo code {\sc noar} (Dumont et al. 2000)
are shown in Fig.~\ref{fig:MonteCarlo}. We see that the slope
of the spectrum at low energies is slightly modified by Compton up-scattering
and the spectrum bends at
high energies, imitating the high energy cut-off. At energies close to 1 MeV
energy losses become lower since the Klein-Nishina effect significantly
decreases the scattering cross-section.

Detailed discussion of possible processes is definitely beyond the
present simple paper. However, in order to see whether any systematic
differences between the 'primary' spectra from the inner and the
middle zone influence objects with different accretion rates we
calculate
what fraction of the total energy dissipated in the corona comes from
the three zones separately.

\begin{figure}
\epsfxsize=8.8cm
\epsfbox{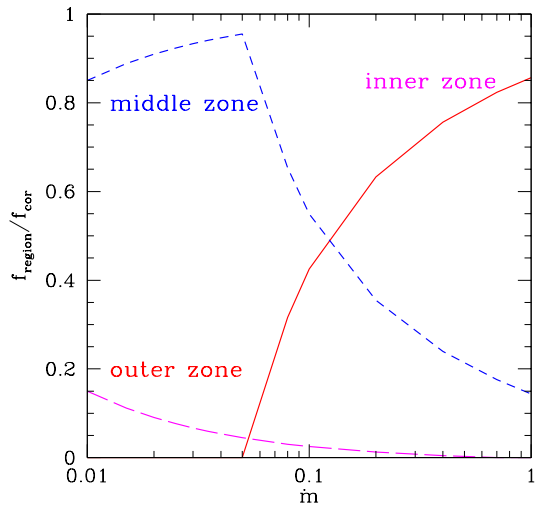}
\caption{The division of the total radiation flux dissipated in the
corona between the three zones as a function of the source accretion rate,
for $f_{cor} = 0.9$. At large
accretion rates the innermost zone (nonthermal emission) always dominates
(continuous lines),
at low accretion rate most of the emission comes from the middle
zone with mostly thermal emission (short-dash lines), and the contribution
 from the outer zone
(long-dashed lines) is small for the adopted  model parameters.}
\label{fig:energy}
\end{figure}

We show an example of such a plot in
Fig.~\ref{fig:energy}. We see that high accretion rate objects,
according to the presented considerations, are dominated by emission
from the inner zone. Therefore, their spectra are expected to be
predominantly non-thermal. Observationally, this should manifest in
the extension of the spectrum possibly well beyond the standard 100
keV. Unfortunately, estimates of the high-energy cut-off of radio
quiet quasars or Narrow Line Seyfert 1 galaxies, believed to accrete
roughly at an Eddington rate, are not available.  In Seyfert 1
galaxies the X-ray emission is predominantly thermal, so if these
sources accrete at a few percent of the Eddington rate their emission
should be
dominated by the middle region. This is indeed the case in our model
if the fraction  of energy dissipated in the corona is 0.65 or larger,
and if the covering factor is of the order of 0.5 or lower.

\subsection{Possible trends in global parameters}

\begin{figure}
\epsfxsize=8.8cm
\epsfbox[50 200 530 680]{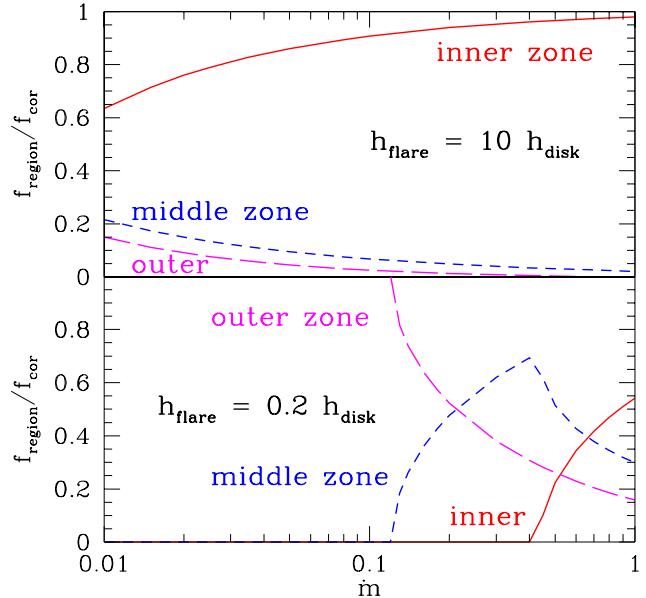}
\caption{The same as Fig.~\ref{fig:energy} but assuming that flares are
located ten times higher above the disk (upper panel). Lower panel shows
the solution appropriate for a magnetically heated disk skin ($f_{cover}=1$,
$h_{flare} = 0.2 h_{disk}$).}
\label{fig:energy2}
\end{figure}

In our model we assumed that the flare is located at the height $2 h$ 
(with $h$ being the cold disk thickness) from the equatorial plane. However,
flares rising to larger distances are also plausible (e.g. Romanova et al. 
1998, di Matteo, Celotti \& Fabian 1999, Poutanen \& Fabian 1999, 
Merloni \& Fabian 2001). This effect would naturally decrease the extension of
the middle zone predicted by our model. We see that in Fig.~\ref{fig:energy2},
upper panel, calculated under assumption that the flare is 
located ten times higher above the disk body. On the other hand, 
high accretion rate
models with magnetic energy transport, like in Merloni (2003) or Czerny et al.
(2003),
may lead to the extended middle zone. 3-D MHD simulations by Turner (2004)
show that most of the dissipation takes place just below the disk surface. 
We can 
model such solution assuming the flare height being only a fraction of the
disk thickness and taking $f_{cover} = 1$ (Fig.~\ref{fig:energy2}, lower
panel).

What is more, a coupling between the model parameters $f_{cor}$, $f_{cover}$
and $\dot m$ can be expected. For example, in a model of Liu et al. 
(2002) the
energy fraction in the corona increases with the radius in the radiation
pressure dominated region and decreases in the gas pressure dominated
region thus peaking somewhere in the transition radius. The model of continuous
accreting corona of \Agata \& Czerny (2000) based on disk evaporation showed
basically an increase of the coronal strength in-wards. The
model of Merloni (2003) predicts two branches of solutions, with the
coronal strength decreasing with the distance at the stable branch. 
Most models 
predict a systematic decrease of
the corona strength with an accretion rate, and such a trend is supported
by observations (Wang, Watarai \& Mineshige 2004).   

Such coupling can be easily incorporated into the present model. The 
computations for the model of
accreting corona (\Agata \& Czerny 2000) show the domination of bare flare 
region in the total energy output since the corona strength decreases 
rapidly with radius, and usually to the presence of an inner hole in the 
cold disk due to the disk evaporation. The radial trend present in the
solution of 
Merloni (2003) decreases the role of the outer region, since the corona
does not form there, while the slight
systematic increase of the coronal strength and its radial extension 
with an increase of the
accretion rate enhances the role of the inner region. Exact predictions 
depend again on the description of the
flare height which is not provided by these models.

\subsection{Characteristic timescales}

All these phenomena happen, however, if the material has enough time to
rise from the disk surface before the flare turns off.

The timescale for the expansion of the hot material with the local sound
speed up to $H_{hot}$ is the
same as the timescale to reach the hydrostatic equilibrium, i.e. it is given by
the Keplerian timescale, as implied by  Eq.~\ref{eq:Hhot}.

The timescale for the formation of a magnetic loop due to magnetorotational
instability and
its emergence from the disk interior is also of the order of the
Keplerian timescale (e.g. Romanova et al. 1998).

The duration of the radiative stage of the flare, however, is under debate.
There are two kinds of argument leading in two directions.

Theoretical arguments are made that the magnetic loop reconnection is
a very fast process, and therefore the duration of the flare is expected to
be much shorter than the formation of the magnetic loop phase (see e.g.
Haardt et al. 1994, Romanova et al. 1998).
This option was also considered by
Nayakshin \& Kazanas (2002) and Collin et al. (2003).

Observationally, the analysis of the AGN variability in the X-ray band leads
to the determination of the power spectra in which most of the power is in
timescales of a day or longer (see e.g. Markowitz et al. 2003,
Uttley et al. 2002) thus corresponding to timescales
longer even by a factor of a few than the Keplerian timescale range in
the disk.

Models explaining the power spectra (mostly in the context of galactic
X-ray sources, but the phenomenon is intrinsically the same apart from the
overall scaling factor due to the central mass) are solving the problem of
observed long timescales by the claim that flares are naturally short lasting
phenomena but they come in groups, or avalanches. The measured power
spectrum reflects basically the coupling timescales between the flares, with
the timescales of single flares contributing only to the steep high frequency
part of the power spectrum (see e.g. Poutanen \& Fabian 1999, \. Zycki 2002).

Direct observational evidence of the development of a single flare is
available only for solar corona flares. High resolution imaging and
spectroscopy on board of the Yohkoh and RHESSI satellites
brought very interesting results.

Magnetic loops emerge rather slowly from the interior to the solar corona,
on a timescale of minutes, so the upward motion of the magnetically dragged
plasma is subsonic. The trigger of the flare - i.e. the hard X-ray emission -
is rather rapid, involving timescales of a few seconds (e.g. Lee et al. 2002a,
Sui \& Holman 2003) and the coronal source forms at the top of the magnetic
loop. This coronal source remains active for a few minutes and then slowly
decreases in brightness; at the same time the entire magnetic loop also
evolves. Such a long active phase seen in the data is not expected from
simple models of the phenomena occurring within current sheets, so some ad hoc
ideas are developed in order to account for the slow release of the magnetic
energy  like for example magnetic trapping (e.g. White et al. 2002). What is
more, high time-resolution observations of 14 behind-the-limb flares,
with shielded
foot-points, show that actually the hard X-ray emission of a single flare,
present for several minutes, consists of many sub-pulses, as if the
reconnection proceeds slowly, in a step by step process (Tomczak 2001).

Therefore, observations of the solar corona rather indicate that the duration
of the flare is of the same order as the evolutionary timescale of the whole
magnetic loop, and not much shorter as implied by the Alfven speed based
estimates. It is clearly an open question whether the same applies to the
flares above an accretion disk in AGN.

\subsection{Time evolution of a flare}

Certain aspects of the time evolution of the flare above an
accretion disk were already
discussed in several papers (e.g. Romanova et al. 1998; di Matteo 1998,
Poutanen \& Fabian 1999, Beloborodov 1999, Lu \& Yu 2001, Czerny et al. 2004).
It was stressed that slow expansion of
the disk atmosphere makes the situation actually time-dependent (not
well represented by pressure equilibrium) and may be responsible for the
observationally confusing issue of the response of the irradiated disk surface
to the change of the continuum
(Vaughan \& Edelson 2001; Nayakshin \& Kazanas 2002, Collin et al. 2003).

The picture which emerges from the results given in
Sect.~\ref{sect:results} is even more
complex than it used to be.

Shortly after the beginning of the flare
activity the irradiated disk surface reaches a new thermal equilibrium.
Later, the heated high-pressure layer
starts to expand in
order to regain the hydrostatic equilibrium, as
described by Collin et al. (2003).

The flare, anchored deeply inside the disk (e.g. Nayakshin \& Kazanas 2001),
is not expected to be influenced
considerably by the rising hot material until this material starts to surround
the flare. This happens after a relatively small fraction of the
Keplerian timescale in the outer zone and relatively late (again, in terms
of the local Keplerian timescale) in the middle zone. The outer zone, with its
flares embedded in an optically thin plasma, resembles actually the solar
corona, with flares inside the extended coronal plasma. Radiation
pressure (not included in computations) will additionally increase the
height of the hot layer although not substantially (gravitational attraction
rising linearly outwards will counter-balance the radiative force as long as
the thickness of the heated zone does not exceed the local disk radius).

The middle zone is
qualitatively different and currently available spectra models
(see Fig.~\ref{fig:MonteCarlo} are not computed
self-consistently. Rising optical depth between the source and the disc's cold
layers must saturate an incident flux. Exact outcome is not clear: for
example, in recent 3-D MHD simulations of the MRI instability localized strong
dissipation takes place at the optical depth of order of $\tau \approx 10^4$
from the disk surface (Turner 2004). It is not clear, however,
if such an extreme optical
depth would be obtained if Comptonization was included as a mean of the
radiative transfer instead of just flux-limited approximation
(i.e. bremsstrahlung/black body). Future modeling should address, in
particular, this issue.

An interesting study of the rise of the disk material under the flare was done
by Williams \& Maletesta (2002) in the context of cataclysmic variables, where
they considered accelerated electrons hitting the disk surface. Some aspects
of this study may also apply to AGN.

\section{Conclusions}

Apart from radiative changes, X-ray flares above AGN accretion disks
also modify the geometry of the disk surface. They cause the material
to expand and, for a certain range of parameters, the flare is
embedded within the disk atmosphere.

The physical changes of the disk atmosphere in the vicinity of a
flare, namely the height $H_{hot}$ of the heated surface and its
optical depth $\tau_{es}$, depend strongly on the radial flare
position $r$. Generally, $H_{hot}$ rises, while $\tau_{es}$ declines
with $r$.

For a certain parameter range of the fraction of coronal heating $f_{cor}$
and of the flare covering factor $f_{cover}$ a zone of the disk exists
where the flare sources are deeply buried within the expanded disk
atmosphere (middle zone). For these flares no typical
Compton reflection component is expected. For example,
for low accretion rate objects with strong corona no relativistically
broadened reflection from the inner part of the disk is predicted by the model.

Active Galaxies accreting at high $\dot m$ rates are dominated by
reprocessed X-rays from the inner zone where the flares are not
surrounded by the expanding disk medium. For these objects we expect
a relativistically broadened Compton reflection from the significantly ionized
surface.

For intermediate accretion rate objects and a specific parameter range, the
model predicts the narrow inner zone so the relativistically broadened
reflection in such sources forms predominantly in an inner, narrow ring.
This is an interesting possibility in the context of the Seyfert galaxy
MCG -6-30-15.

\section*{Acknowledgments}

We thank Suzy Collin and Martine Mouchet for very helpful discussions
and comments to the manuscript.
Part of this work was supported by the grants 2P03D00322 and PBZ-KBN-054/P03/2001
of the Polish
State Committee for Scientific Research, by the
Laboratoire Europe\' en Associ\' e Astrophysique Pologne-France, and by
the Hans-B\"ockler-Stiftung.

\ \\
This paper has been processed by the authors using the Blackwell
Scientific Publications \LaTeX\  style file.

\end{document}